\documentclass[english,showpacs,floatfix,12pt]{revtex4}
\usepackage[dvips]{graphicx}
\usepackage{longtable}
\usepackage{amsmath,amssymb}
\usepackage{dcolumn}
\usepackage{latexsym}
\usepackage{babel}
\usepackage[latin1]{inputenc}
\usepackage{color}
\usepackage[colorlinks]{hyperref}
\def\pa{\partial}
\def\al{\alpha}
\def\ga{\gamma}
\def\dl{\delta}
\def\Dl{\Delta}
\def\la{\lambda}
\def\be{\beta}
\def\kp{\kappa}
\def\th{\theta}
\def\sg{\sigma}
\def\vf{\varphi}
\def\nb{\nabla}
\def\lan{\langle}

\def\l{\left}
\def\r{\right}
\def\nn{\nonumber}
\def\ck{\check}
\def\diag{\mbox {diag}}
\def\E{{\cal{E}}}
\def\wt{\widetilde}

\begin{document}
\title{The Characteristic Functions and Their Typical Values
\\ for the Nonlinear Spinors}
\author{Ying-Qiu Gu}
\email{yqgu@fudan.edu.cn} \affiliation{School of Mathematical
Science, Fudan University, Shanghai 200433, China} \pacs{29.90.+r,
11.10.Ef, 31.15.aj, 12.20.Fv}
\date{6th June 2009}

\begin{abstract}
In this paper, we solve the eigen solutions to some nonlinear
spinor equations, and compute several functions reflecting their
characteristics. The numerical results show that, the nonlinear
spinor equation has only finite meaningful eigen solutions, which
have positive discrete mass spectra and anomalous magnetic moment.
The nonlinear potential and interactions yield different
contributions to the total energy, and these components of the
energy lead to different energy-speed relation. The magnitude of
these components can be detected by elaborate experiments. The
weird properties of the nonlinear spinors might be closely related
with the elementary particles and their interactions, so some
deeper investigations on them are significant.\vskip3mm
\noindent{{Keywords}}: {\sl nonlinear spinor field, anomalous
magnetic moment, mass-energy relation}
\end{abstract}
\maketitle

\section{Introduction}
\setcounter{equation}{0}

Since Dirac established relativistic quantum mechanics, many
scientists such as H. Weyl, W. Heisenberg, have attempted to
associate the elementary particles with the eigenstates of the
nonlinear spinor equation\cite{1,2,3,4,5,6}. In 1951, R.
Finkelsten solved some rigorous solutions of the nonlinear spinor
equation by numerical simulation, and pointed out that the
corresponding particles have quantized mass spectra\cite{7,8}.
However these researches have not realized their authors' goal due
to the mathematical difficulties in analyzing nonlinear spinor
equation.

In recent years, a great effort has been made along this line of
research. The theoretical proof about the existence of solitons
was investigated in \cite{12,13,14,15,16,17}. The symmetries and
many conditional exact solutions of the nonlinear spinor equations
are collected in \cite{exact}. The present work is a development
of some previous works\cite{gu1,gu2,gu3}. In this paper, we define
some functions which reflect the properties of eigen solutions to
the nonlinear spinor equations, and compute the typical values,
then extract some important information from the data. The
following are some general knowledge for the nonlinear spinors.

Denote the Minkowski metric by $\eta_{\mu\nu}={\rm
diag}(1,-1,-1,-1)$, Pauli matrices by
\begin{eqnarray}
 {\vec\sg}=(\sg^{j})= \l \{\l (\begin{array}{cc}
 0 & 1 \\ 1 & 0 \end{array} \r),\l (\begin{array}{cc}
 0 & -i \\ i & 0 \end{array} \r),\l (\begin{array}{cc}
 1 & 0 \\ 0 & -1 \end{array} \r)
 \r\}.\label{1.1}\end{eqnarray}
Define $4\times4$ Hermitian matrices as follows
\begin{eqnarray}\al^\mu=\l\{\l ( \begin{array}{cc} I & 0 \\
0 & I \end{array} \r),\l (\begin{array}{ll} 0 & \vec\sg \\
\vec\sg & 0 \end{array}
\r)\r\},\quad \ga =\l ( \begin{array}{cc} I & 0 \\
0 & -I \end{array} \r),
\quad \be=\l (\begin{array}{cc} 0 & -iI \\
iI & 0 \end{array} \r),\label{1.2}
\end{eqnarray}
where $\mu\in\{0,1,2,3\}$, $x^0=ct$ and $\al^\mu=\ga^0\ga^\mu$. In
this paper, we adopt the Hermitian matrices (\ref{1.2}) instead of
Dirac matrices $\ga^\mu$ for the convenience of calculation. For
Dirac's bispinor $\phi$, the quadratic forms of $\phi$ are defined
by
\begin{eqnarray}
  \ck\al^\mu=\phi^{+}\al^\mu\phi,\qquad \ck\ga=\phi^{+}\ga \phi,
\qquad \ck\be=\phi^{+}\be \phi,
  \label{1.3}  \end{eqnarray}
where the superscript `+' stands for the transposed conjugation.
By the definition (\ref{1.3}) we have $\ck\al^\mu=\phi^\dag
\ga^\mu \phi$ etc., where $\phi^\dag=\phi^+\ga^0$ is the Dirac
conjugation\cite{22}. By transformation law of $\phi$, one can
easily check that $\ck\al^\mu$ is a contra-variant 4-vector,
$\ck\ga$ a true scalar and $\ck\be$ a pseudo-scalar. One can
construct some other covariant quadratic forms, but they are not
independent on (\ref{1.3}) for some Pauli-Fierz
identities\cite{gu1,20,21}, such as
$\ck\al_\mu\ck\al^\mu=\ck\ga^2+\ck\be^2$.

In general, the Lagrangian of the nonlinear bispinor $\phi$ with a
vector potential $A^\mu$ and scalar $G$ is given
by\cite{Tgu1,Tgu2}
\begin{eqnarray}
{\cal L}&=&\phi^+\al^\mu (i\pa_\mu-e A_{\mu})\phi-\mu\ck
\ga+V(\ck\ga,\ck\be)-s\ck\ga G \nn
\\ &~&-\frac 1 2 \kp(\pa_\mu A_\nu\pa^\mu A^\nu-a^2 A_\mu A^\mu)-\frac 1 2
\la (\pa_\mu G\pa^\mu G-b^2 G^2), \label{lag}
\end{eqnarray} where $A^\mu$ and $G$ include the self and
external potential, $\kp=\pm 1$ and $\la=\pm 1$ are used to stand
for the repulsive or attractive self interaction, e.g. $A^\mu$
stands for repulsive electromagnetic potential if $(\kp=1, a=0)$,
but stands for attractive interactive potential similar to strong
interaction if $(\kp=-1,a\ne 0)$. $G$ stands for a scalar
interactive potential like Higgs field, which is repulsive if $\la
=1$, and attractive if $\la=-1$. However, in this paper, we take
(\ref{lag}) as one system, and only the internal interactions are
considered.

If $\pa_{\ck\be}V \ne 0$, the eigen solution might be absent, so
we only consider the case $V=V(\ck\ga)>0$ is a concave function
satisfying
\begin{eqnarray} V'(\ck\ga)\ck\ga>V(\ck\ga),\qquad {\mbox{if}}~~
(\ck\ga>0). \end{eqnarray} The corresponding dynamical equation is
given by
\begin{eqnarray}
\al^\mu(\hbar i\pa_\mu-e A_\mu)\phi &=&(\mu c+s G-V')\ga\phi, \label{1.04} \\
(\pa_\al\pa^\al +a^2) A^\mu &=&\kp e\ck\al^\mu,
\label{1.5}\\
(\pa_\al\pa^\al +b^2) G~&=&\la s\ck\ga. \label{1.05}
\end{eqnarray}
The Hamiltonian form of (\ref{1.04}) is given by
\begin{eqnarray}\hbar i \pa_t\phi=\hat H\phi,\quad \hat H=c[eA_0+\vec\al\cdot\hat p +(\mu c+s G-V')\ga]. \label{1.4}
\end{eqnarray} where $\hat
p=-\hbar i\nb+e\vec A$ is the momentum operator. For (\ref{1.4}),
by the current conservation law, we have the normalizing condition
\begin{eqnarray}\int_{R^3}|\phi|^2d^3x= 1.  \label{1.6}
\end{eqnarray}

Let $\hat J$ be the angular momentum operator
\begin{eqnarray}
\hat J=\vec r \times \hat p+\frac 1 2 \hbar \vec \ga,\qquad
\ga_k=\diag (\sg_k,\sg_k), \label{2.2}
\end{eqnarray}
then the eigenfunctions of $\hat J_3=-\hbar i \pa_\vf+\frac 1 2
\hbar\ga_3$ are given by
\begin{eqnarray}
\hat J_3 \phi_j=j_3\hbar\phi_j,\qquad \phi_j=(u_1,u_2e^{\vf
i},iv_1, iv_2e^{\vf i})^T e^{j \vf i}, \label{1.12}
\end{eqnarray}
where the index `T' stands for transpose, $j_3=j+\frac 1 2,
j\in\{0,\pm 1,\pm 2,\cdots\}$. For all the eigenfunctions, $\hat
J_3$ is commutative with the nonlinear Hamilton operator like the
linear case, so under a suitable choice of coordinate system, the
solutions of (\ref{1.4}) take the following
form\cite{commu,eigen},
\begin{eqnarray}
\phi_j=(u_1,u_2e^{\vf i},iv_1,iv_2e^{\vf i})^T \exp(j \vf
i-\frac{mc^2} \hbar it),
\end{eqnarray}
where $u_k,v_k(k=1,2)$ are real functions independent of $\vf$ and
$t$.

The system (\ref{1.4}) has many symmetries such as the global
gauge invariance, the spatial reversal invariance etc. The above
assumptions have removed the uncertainty caused by the symmetry,
and the solutions are determined except for a signature. This
procedure is quite important for the numerical solving and
stability analysis. If $V(\ck\ga)$ take the form of polynomials,
the solutions are analytic functions of $r$, and then they can be
expressed as the Taylor series of $r$. In the cases of $j_3 =\pm
\frac 1 2$, which are the only cases for free particles, we have
the formal solution as follows
\begin{eqnarray}
u_1+u_2 i&=&\sum _{m=0}^{\infty } {r}^{2 m} \left( K_{{m}}{e^{-2 i
m\th}}+ \sum _{n=-m+1}^{m}A_{mn}{e^{2 i n\th}}
\right),\label{1.14} \\
v_1+v_2 i&=&\sum _{m=0}^{\infty } {r}^{2 m+1} \left( J_{{m}}{e^{-i
 \left( 2 m+1 \right) \th }}+\sum _{n=-m}^{m}B_{mn}{e^{i \left( 2 n+
1 \right) \th }} \right), \label{1.15}
\end{eqnarray}
where $K_m,J_m$ are real free parameters determined by boundary
conditions and the normalizing condition, but $(A_{mn},B_{mn})$
are real numbers determined by $(K_n,J_n)$ with $n \le m$. The
normalizing condition (\ref{1.6}) becomes
\begin{eqnarray}
2\pi\int^\infty_0 r^2 dr\int^\pi_0\sin\th d\th
(u_1^2+u_2^2+v_1^2+v_2^2)=1. \label{2.7}
\end{eqnarray}

\section{properties of the dark nonlinear spinor}
\setcounter{equation}{0}

The simplest case of (\ref{1.4}) is the following dynamical
equation
\begin{eqnarray}
\hbar i \pa_t\phi =\hat H \phi,\qquad \hat H=c[\vec\al\cdot\hat
p+(\mu c -w\ck\ga )\ga].\label{2.1}
\end{eqnarray}
Different from the linear case, the nonlinear spinor equation
generally has continuous spectra if the restriction of (\ref{1.6})
is absent, so the normalizing condition is a quantizing condition
for nonlinear spinors, and the nonlinear coupling coefficient $w$
is meaningful only if the solution satisfies the normalizing
condition (\ref{1.6}).

The eigen solutions to (\ref{2.1}) with spin $j_3 =\pm \frac 1 2$
can be solved rigorously as follows
\begin{eqnarray}\l\{\begin{array}{ll}
\phi_{e\uparrow}=(g,0,if\cos\th,if\sin \th e^{\vf i})^{T}\exp
  (-i\frac{mc^2}{\hbar}t),& {\mbox {for~}} (\textsf{P}=1,~j_3  =\frac 1 2)\\
\phi_{e\downarrow}=(0,g,if\sin \th e^{-\vf i},-if\cos\th)^{T}\exp
  (-i\frac{mc^2}{\hbar}t),& {\mbox {for~}} (\textsf{P}=1,~j_3  =-\frac 1 2)\\
 \phi_{o\uparrow}=(f\cos\th,f\sin \th e^{\vf i},ig,0)^{T}\exp
  (-i\frac{mc^2}{\hbar}t),& {\mbox {for~}} (\textsf{P}=-1,~j_3  =\frac 1 2)\\
  \phi_{o\downarrow}=(f\sin \th e^{-\vf i},-f\cos\th,0,ig)^{T}\exp
  (-i\frac{mc^2}{\hbar}t),& {\mbox {for~}} (\textsf{P}=-1,~j_3  =-\frac 1 2)\end{array}\r.
  \label{2.8}  \end{eqnarray}
where $\textsf{P}=1$ corresponds to even parity, and
$\textsf{P}=-1$ corresponds to odd parity. For the above
eigenfunctions, we have
\begin{eqnarray}
  \ck\ga=\textsf{P}(g^2-f^2),\qquad   4\pi\int^\infty_0(g^2+f^2)r^2dr=1.
\label{2.9}  \end{eqnarray} The radial equation of even parity
satisfies
\begin{eqnarray}
  \l\{ \begin{array}{l} \frac d {dr} g=-\frac{1}{\hbar c}[(\mu+m)c^2-wc(g^2-f^2)]f,\\
\frac d {dr} f=-\frac{1}{\hbar
c}[(\mu-m)c^2-wc(g^2-f^2)]g-\frac{2}{r}f.
  \end{array}\r.  \label{2.10}
\end{eqnarray}
For the odd parity, we have
\begin{eqnarray}
  \l\{ \begin{array}{l} \frac {d}{dr}g=-\frac{1}{\hbar c}[(\mu-m)c^2+wc(g^2-f^2)]f,\\
   \frac {d}{dr}f=-\frac{1}{\hbar c}[(\mu+m)c^2+wc(g^2-f^2)]g-\frac{2}{r}f.
  \end{array}\r.  \label{2.11}
\end{eqnarray}
The initial data of (\ref{2.10}) and (\ref{2.11}) satisfy
$f(0)=0,~ g(0)>0$. For (\ref{2.10}) and (\ref{2.11}), we have
positive mass $0< m<\mu$ if and only if $w> 0$\cite{gu2}.

Making transformation
\begin{eqnarray}
a&=&\sqrt{\frac{\mu+m}{\mu-m}},\qquad
r_0=\frac{\hbar}{c\sqrt{\mu^2-m^2}}
  =\frac{(a^2+1)\hbar}{2a\mu c}, \qquad \rho = \frac{r}{r_0},
  \label{2.12} \\
u &=& \sqrt{\frac{w(a^2+1)}{2a\mu c}}g, \qquad
v=-\sqrt{\frac{w(a^2+1)}{2a\mu c}}f.   \label{2.13}
\end{eqnarray}
where $a$ is equivalent to the spectrum, $r_0$ takes the unit of
length. (\ref{2.10}) and (\ref{2.11}) can be rewritten in a
dimensionless form. For (\ref{2.10}) we have
\begin{eqnarray}
  \l\{\begin{array}{ll} u'=(a-u^2+v^2)v,\qquad & u(0)=u_0> 0,\\
       v'=(\frac 1 a-u^2+v^2)u-\frac{2}{\rho}v,\qquad & v(0)=0, \end{array}\r.
       \label{2.14}   \end{eqnarray}
where prime stands for $\frac d{d\rho}$. For (\ref{2.11}) we have
\begin{eqnarray}
  \l\{\begin{array}{ll} u'=(\frac 1 a+u^2-v^2)v,\qquad & u(0)=u_0> 0,\\
       v'=(a+u^2-v^2)u-\frac{2}{\rho}v,\qquad & v(0)=0, \end{array}\r.
       \label{2.15} \end{eqnarray}
The normalizing condition (\ref{2.9}) becomes
\begin{eqnarray}
  \l( a+a^{-1}\r)^2{\int^\infty_0(u^2+v^2)\rho^2 d\rho}=S^2 \equiv {\frac{w\mu^2 c^2}{\pi\hbar^3}},
  \label{2.16}  \end{eqnarray}
where $S$ is a dimensionless constant to be determined.

The computation shows that, for any given $a>1$, there exists a
sequence of initial data $0<u(0)_1<u(0)_2<\cdots$, such that
(\ref{2.10}) and (\ref{2.11}) have eigen solutions. The
theoretical analysis proves that there are infinite eigen
solutions for every $a$\cite{14}. In \cite{gu2} we have shown
three families of eigen solutions with even parity and the first
family of eigen functions with odd parity.

To describe the properties of the eigen solutions, we define the
following dimensionless functions, which are continuous functions
of spectrum $a$ for the same family solutions.
\begin{enumerate}

\item The dimensionless norm $y(a)$
\begin{eqnarray}
y\equiv\frac 1 2 {\lg}\l((a+a^{-1})^2{\int^\infty_0(u^2+v^2)\rho^2
d\rho}\r). \label{2.17}
\end{eqnarray} For the same family of eigen solution, $y$ is a
continuous function of $a$. By (\ref{2.16}), the normalizing
condition is equivalent to the equation $y = \lg S $.

\item The dimensionless energy $\E(a)$
\begin{eqnarray}
\E &\equiv& \frac 1 {\mu c^2}\l(mc^2+\frac{1}{2}wc\int^\infty_0
\ck\ga^2\cdot 4\pi r^2
dr\r)\nn\\
&=& \frac{a^2-1}{a^2+1}+\frac{a}{a^2+1}\frac{\int^\infty_0
(u^2-v^2)^2\rho^2 d\rho}{\int^\infty_0(u^2+v^2)\rho^2 d\rho}.
\label{2.18}  \end{eqnarray} This definition of energy is in the
N\"other's sense.

\item The mean diameter of an eigen solution $d(a)$
\begin{eqnarray}
d\equiv \frac 2 \la \frac{\int r|\phi|^2 d^3 x}{\int |\phi|^2 d^3
x}=\frac{a^2+1}a \frac{\int_0^\infty(u^2+v^2)\rho^3
d\rho}{\int_0^\infty(u^2+v^2)\rho^2 d\rho}, \label{2.19}
\end{eqnarray}
where $\la=\frac \hbar{\mu c}$ is a universal Compton wave length
for all solutions.

\item The total dimensionless inner pressure $P(a)$
\begin{eqnarray}
P&\equiv&\frac 1 {3\mu c^2}\l(mc^2-\int^\infty_0 (\mu c^2
\ck\ga+\frac1 2 wc\ck\ga^2)\cdot 4\pi r^2 dr\r)\nn \\
&=&\frac 1 3 \l(\frac{a^2-1}{a^2+1}-\frac{\int^\infty_0
(u^2-v^2)\rho^2 d\rho}{\int^\infty_0(u^2+v^2)\rho^2
d\rho}-\frac{a}{a^2+1}\frac{\int^\infty_0 (u^2-v^2)^2\rho^2
d\rho}{\int^\infty_0(u^2+v^2)\rho^2 d\rho}\r). \label{2.20}
\end{eqnarray}
 \end{enumerate}

The physical meanings of $y(a)$, $\E(a)$ and $d(a)$ are evident.
Now we examine the meanings of $P(a)$. For the perfect fluid in
relativity, the energy momentum tensor is given by
\begin{eqnarray}
T^{\mu\nu}=(\rho_m+P)U^\mu U^\nu-P g^{\mu\nu},\qquad
T^\mu_\mu=\rho_m-3 P. \label{2.22} \end{eqnarray} For the static
fluid, we have the 4-dimensional speed $U^\mu=(1,0,0,0)$, and then
\begin{eqnarray}
T^0_0=\rho_m,\qquad P=\frac 1 3 (T^0_0-T^\mu_\mu). \label{2.23}
\end{eqnarray}

For the nonlinear spinor (\ref{2.1}) in curved space-time with
diagonal metric, we define the corresponding concepts as
follows\cite{gu4,gu5,gu6}
\begin{eqnarray}\begin{array}{lll}
T^{\mu\nu}&=&\frac 1 2 \Re\lan\phi^+(\varrho^\mu
i\pa^\nu+\varrho^\nu i\pa^\mu)\phi\rangle-{\cal L} g^{\mu\nu}\\
&=&\frac 1 2 \Re\lan\phi^+(\varrho^\mu i\pa^\nu+\varrho^\nu
i\pa^\mu)\phi\rangle+(V'\ck\ga-V)g^{\mu\nu}.\end{array}
\label{2.24} \end{eqnarray} For static spinor, we have
\begin{eqnarray}
P = \frac 1 3 (T^0_0-T^\mu_\mu)=\frac 1 3 (m
|\phi|^2-\mu\ck\ga-\frac 1 2 w\ck\ga^2). \label{2.26}
\end{eqnarray}
The dimensionless form of the total inner pressure of the spinor
becomes (\ref{2.20}).

\begin{figure}[ht]
\centering
\includegraphics[width=12cm]{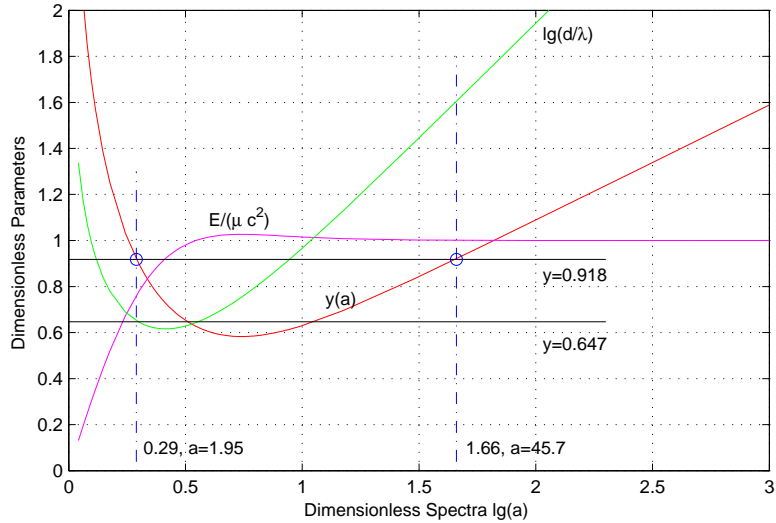}
\caption{The norm function $y(a)$, dimensionless energy $\E=\frac
E {\mu c^2}$ and mean diameter $d(a)$ of a spinor. Only the
solutions corresponding to the intersection $y(a)=\lg S$ are
meaningful in physics} \label{fig1}
\end{figure}

\begin{figure}[ht]
\centering
\includegraphics[width=12cm]{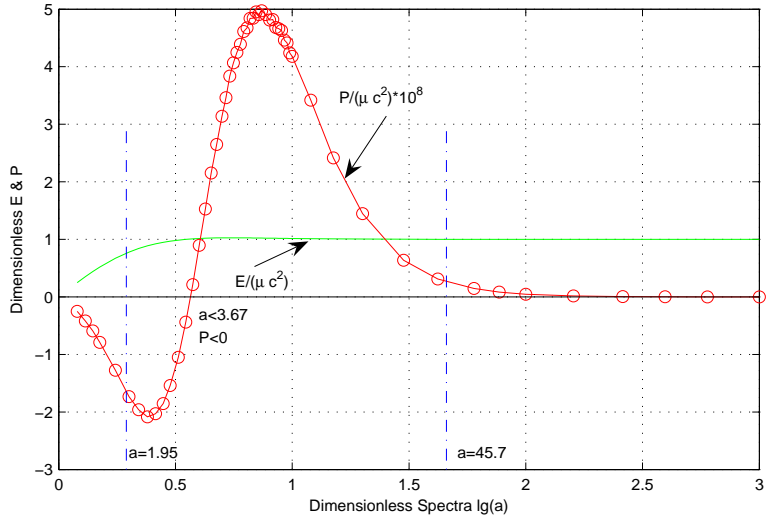}
\caption{The total energy $\E(a)$ and inner pressure $P(a)$ of a
dark spinor} \label{fig2}
\end{figure}

The curves of the dimensionless functions defined above are shown
in Fig.\ref{fig1} and Fig.\ref{fig2}. In Fig.\ref{fig1}, the
normalizing condition $y\equiv \lg S=0.918$ and $y\equiv \lg
S=0.647$ are derived from the anomalous magnetic moment(AMM) of an
electron according to different definition of mass, as computed in
the next section. A rough computation was once given in
\cite{gu3}.

For an electron, we have $\mu\dot= m_e=9.11\times 10^{-21}$kg,
$\hbar=1.055\times 10^{-34}$J.s, $c=2.998\times 10^8$m/s. By
(\ref{2.16}) and $S=8.277$, we can estimate the value
\begin{eqnarray}
w=\frac{\pi\hbar^3S^2}{\mu^2c^2}\dot=4.945\times 10^{-59}
S^2=3.385\times 10^{-57}{\rm (J s m^2)}.\end{eqnarray} In this
case, the nonlinear spinor equation has only two valid eigen
solutions corresponding to $a=1.95$ and $a=45.7$. The norm
function $y(a)$ of all other families of eigen solutions have no
intersection points with $y=0.918$.
\begin{figure}[ht]
\centering
\includegraphics[width=12cm]{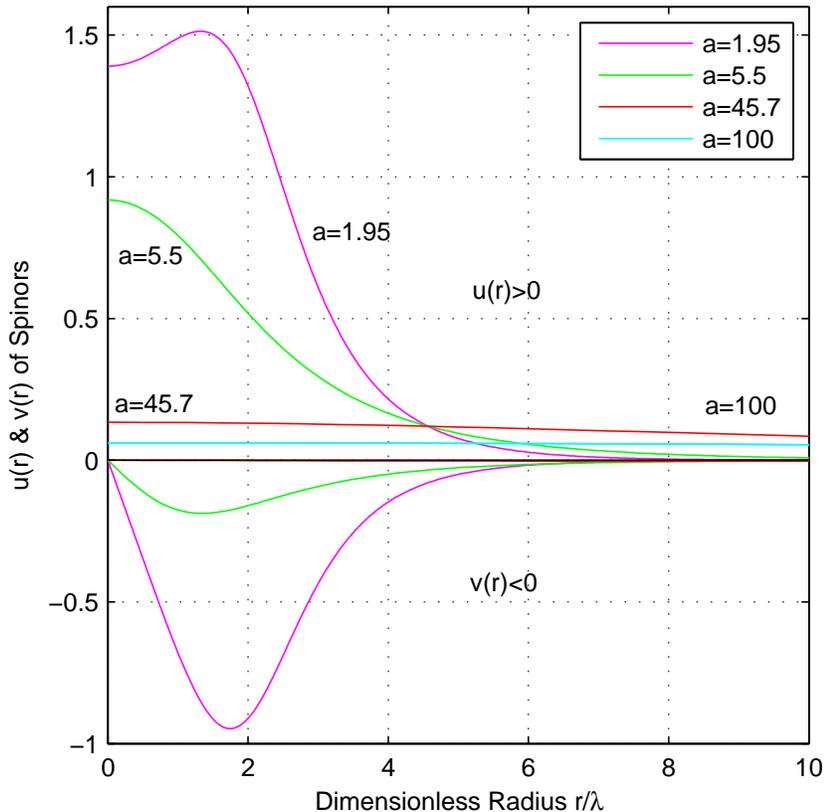}
\caption{The radial distribution of the nonlinear dark spinors}
\label{fig3}
\end{figure}

The radial functions $(G,F)$ of solutions with even parity are
shown in Fig.\ref{fig3}, where
\begin{eqnarray}
G(r)=\sqrt{\frac w {2\mu c}}g=\sqrt{\frac a {a^2+1}}u,\qquad
F(r)=-\sqrt{\frac w {2\mu c}}f=\sqrt{\frac a
{a^2+1}}v.\end{eqnarray} The unit of the coordinate $r$ is the
universal Compton wave length $\frac \hbar {\mu c}$. So the images
of different solutions are visually comparable in Fig.\ref{fig3}.
More images see \cite{gu2}.

\section{The nonlinear spinor with electromagnetic interaction}
\setcounter{equation}{0}

The nonlinear spinor with self electromagnetic interaction was
researched by a few authors. In 1966, M. Wakano has approximately
analyzed the cases of $A_0$ dominance and $\vec A$ dominance when
$w=0$, and reached the following conclusions\cite{[4]}. In the
case of $\vec A$ dominance, the eigen solutions or the solitons do
not exist for the first order approximation. In the case of $A_0$
dominance, the eigen solutions exist but all with negative energy.
In fact, the negative mass is equivalent to change the sign of
$A_0$, which implies to transform the repulsive potential of $A_0$
into the absorbent one. M. Soler and A. F. Ra\~nada calculated the
eigen solutions of (\ref{1.4}) by omitting $\vec A$. But they
neglected the normalizing condition and did not use the true value
of $e$. Their explanation for the results seems to be misguided by
some inadequate ideas\cite{[5],[6]}. Besides, the eigen solutions
with Born-Infeld potential were studied in \cite{[7]}. The
detailed non-relativistic approximation of the many-spinors
equations was given in \cite{gu2,gu7}

In general, the coordinates $r$ and $\th$ can not be separable for
nonlinear spinor with vector potential due to the term $\vec A$.
However $u_k$ and $v_k$ can be conveniently expressed by Fourier
series with respect to $\th$ as (\ref{1.14}) and (\ref{1.15}), and
the equations of the radial functions can be derived via variation
principle, because the eigen solutions are the critical points of
the following energy functional
\begin{eqnarray}
J=2\pi \int^\infty_0 r^2dr\int^\pi_0\sin \th d\th(\phi^+\hat
H\phi+ \frac 1 2 w c \ga^2-\frac 1 2 c\nb A_\mu \cdot\nb
A^\mu-mc^2 \ck\al _0)+mc^2. \label{1.7}
\end{eqnarray}
So the problem (\ref{1.4}) can be changed into an ordinary
differential equation system which can be solved by numerical
computation.

In this paper, we only consider the eigen solutions with $\frac 1
2$-spin and even parity, which is the only valid case for a free
electron. In the dimensionless form, we have the magnitude for the
fields
\begin{eqnarray}
|\vec A|\sim \frac \al a |g|,\quad |A_0|\sim \al |g|,\quad |f|\sim
\frac 1 a |g|,\quad \al\dot = \frac 1 {137},\label{*3.2}
\end{eqnarray}
where $a$ is the dimensionless spectrum. Since the high order
terms are caused by the vector potential $|\vec A| \sim \frac \al
a |g|$, for adequately large $a$, we only keep the first order
approximation for simplicity. Then we have
\begin{eqnarray}
\phi\dot =(g,0,if\cos\th,if\sin\th e^{\vf i})^T \exp(-i\frac{mc^2}
\hbar t), \label{3.1} \end{eqnarray} where $g$ and $f$ are real
functions of $r$ with $g(0)>0$. For large spectrum $a=49.12$, the
relative error of the approximation is less than $10^{-4}$, so the
approximation is accurate enough to reveal the anomalous magnetic
moment of a spinor with electromagnetic field. The less the value
of $a$, the large the error of approximation.

The quadratic forms of $\phi$ are given by
\begin{eqnarray}
\ck\al _0=g^2+f^2,\qquad \ga=g^2-f^2,\qquad \ck {\vec\al}
=2gf\sin\th(-\sin\vf,\cos\vf,0). \label{3.2}
\end{eqnarray} Correspondingly we have
\begin{eqnarray}
A_0=A_0(r),\qquad \vec A=A(r)\sin\th(-\sin\vf,\cos\vf,0).
\label{3.3}
\end{eqnarray} Substituting (\ref{3.2}), (\ref{3.3}) into (\ref{1.7}) we
get the energy functional
\begin{eqnarray}
J &\dot=& 4\pi c\int^\infty_0 r^2dr\l\{\hbar[(f'+\frac 2 r
f)g-g'f]+(\mu-m)cg^2 -(\mu+m)cf^2-\frac 1 2w(g^2-f^2)^2 \r. + \nn \\
 & & \l. e(g^2+f^2)A_0-\frac 4 3 egf A+\frac 1 2 A_0(\pa_r^2+\frac 2
r\pa_r)A_0- \frac 1 3 A(\pa_r^2+\frac 2 r \pa_r-\frac 2
{r^2})A\r\}+mc^2.~~ \label{3.4}
\end{eqnarray} The approximation is only caused by the vector potential $\vec A$.
By variation, we get a closed system of ordinary differential
equations
\begin{eqnarray}
  \l\{ \begin{array}{l} g'=-\frac{1}{\hbar}[(\mu+m)c-eA_0-w(g^2-f^2)]f-\frac 2 {3\hbar}eA g,\\
  f'=-\frac{1}{\hbar}[(\mu-m)c+eA_0-w(g^2-f^2)]g+(\frac 2 {3\hbar}eA-\frac{2}{r})f,\\
  A_0''+\frac 2 r A_0'=-e(g^2+f^2),\quad A''+\frac 2 r A'-\frac 2 {r^2} A=-2egf.\\
  \end{array}\r.  \label{3.5}
\end{eqnarray}

Make transformation
\begin{eqnarray}
&&r_0=\frac \hbar{\sqrt{\mu^2-m^2}c},
~~~a=\sqrt{\frac{\mu+m}{\mu-m}},
~~~\al=\frac{e^2}{4\pi\hbar}=\frac 1 {137.035999},
\label{3.6}\\
&&\rho=\frac r{r_0},
~~u=\sqrt{\frac{wr_0}\hbar}g,~~v=-\sqrt{\frac{wr_0}\hbar}f,~~
  P =\frac{er_0}\hbar A_0,~~Q=\frac{2er_0}{3\hbar}A,
\label{3.7} \end{eqnarray} where $P$ is dimensionless potential,
which can not be confused with the pressure defined in
(\ref{2.20}). Substituting them into (\ref{3.5}), we get the
dimensionless form
\begin{eqnarray}
\l\{ \begin{array}{l}
  u'=(a-P -u^2+v^2)v-Q u,\\
  v'=(\frac 1 a +P -u^2+v^2)u+(Q-\frac 2 \rho)v,\\
  P ''+\frac 2 \rho P '=-\al\frac{u^2+v^2}{\int^\infty_0(u^2+v^2)\rho^2 d\rho},\\
  Q''+\frac 2 \rho Q'-\frac 2{\rho^2}Q=\frac{4\al}{3}\frac{uv}{\int^\infty_0(u^2+v^2)\rho^2 d\rho},
 \end{array}\r. \label{3.8} \end{eqnarray}
In (\ref{3.8}) only $a$ is a free parameter, which acts as the
spectrum similar to the dark case of $e=0$. The normalizing
condition is still (\ref{2.16}). (\ref{3.8}) is independent on the
undetermined coefficient $w$, but it becomes a global problem. The
natural boundary conditions are given by
\begin{eqnarray}\l\{ \begin{array}{l}
u(0)>0,~~v(0)=P '(0)=Q(0)=Q'(0)=0,\\
u\to u_\infty e^{-\rho},~~v\to\frac {u_\infty} a e^{-\rho},~~P \to
\frac \al {4\pi\rho},~~Q\to \frac
{Q_\infty}{\rho^2},~~(\rho\to\infty).
  \end{array}\r.  \label{3.9} \end{eqnarray}

The solutions of $(P,Q)$ can be expressed as
\begin{eqnarray}
P &=& \frac \al{\int^\infty_0(u^2+v^2)\rho^2d\rho}
{\int_\rho^\infty \frac 1 {\rho^{2}}\int_0^\rho
\l[u^2(\tau)+v^2(\tau)\r]\tau^2d\tau
d\rho},   \label{3.10} \\
Q &=& \frac {-4\al}{3\rho^2\int^\infty_0(u^2+v^2)\rho^2d\rho}
{\int_0^\rho \rho^2 \int_\rho^\infty u(\tau)v(\tau)d\tau  d\rho}.
  \label{3.11} \end{eqnarray}
We have $P>0,Q>0$ for the meaningful solutions. The solution of
(\ref{3.10}) and (\ref{3.11}) can be soundly solved by iterative
algorithm.

The total energy of the system in N\"other's sense is given by
\begin{eqnarray}
E &\equiv& \int_{R^3}c\left(\sum_{\forall f} \frac{\pa {\cal
L}}{\pa(\pa_t f)}\pa_t f-{\cal L}\right)d^3 x \nn\\
&=& 2\pi \int^\infty_0 r^2dr\int^\pi_0\sin \th d\th(\phi^+\hat
H\phi+ \frac 1 2 wc \ga^2-\frac 1 2 c\nb A_\mu \cdot\nb A^\mu),
\label{3.12}
\end{eqnarray}
Substituting (\ref{3.6}), (\ref{3.7}) into it, we get the
dimensionless form
\begin{eqnarray}
\E =\frac E{\mu c^2}=
\frac{a^2-1}{a^2+1}+\frac{a}{a^2+1}\frac{\int^\infty_0
[(u^2-v^2)^2 - P (u^2+v^2)-2 Q u v]\rho^2
d\rho}{\int^\infty_0(u^2+v^2)\rho^2 d\rho}.
  \label{3.13} \end{eqnarray}

The mass of a particle is a complex classical concept, which
depends on the method of measurement and the context of theory.
Using different definition of mass, we will get different spectrum
$a$ and constant $S$. Different contribution to the energy has
different energy-speed relation, which can be detected by
elaborate experiment\cite{Tgu2}. Such experiment might be a key to
disclose the structure of elementary particles. We give some more
discussions in the next section.

In what follows, we take $m_{\rm e}$ and $\mu$ as the classical
mass for computation. To get the anomalous magnetic moment, we
introduce an infinitesimal external magnetic field
\begin{eqnarray}
\vec B_{ext}=(0,0,B),~~\vec A_{ext}=\frac 1 2 B(-y,x,0)=\frac 1 2
Br\sin \th(-\sin\vf, \cos\vf,0).
  \label{3.14} \end{eqnarray}
Adding $\vec A_{ext}$ to (\ref{3.3}) and substituting it into
(\ref{3.4}), we get the increment of the energy
\begin{eqnarray}
\Dl E=|\frac {8\pi} 3 ec\int^\infty_0 gf r^3 dr| B\equiv \mu_z B,
 \label{3.15} \end{eqnarray} where $\mu_z$ is
the magnetic moment of the spinor. The dimensionless form is given
by
\begin{eqnarray}
\mu_z=\frac {2(a^2+1)}{3a}\frac{k |\int^\infty_0 uv\rho^3
d\rho|}{\int^\infty_0 (u^2+v^2)\rho^2 d\rho}\cdot \mu_B,\qquad
\mu_B\equiv \frac{e\hbar}{2m_{k}}, \label{3.16} \end{eqnarray}
where the constant $\mu_B$ is the Bohr magneton,
\begin{eqnarray}
k=\l\{\begin{array}{lll} 1 &{\rm ~if~~} m_k=\mu,\\ \E &{\rm ~if~~}
m_k=m_{\rm e}.\end{array} \r.
\end{eqnarray}

By (\ref{3.16}), we get the anomalous magnetic moment of a
particle
\begin{eqnarray}
\Dl g \equiv \frac{\mu_z-\mu_B}{\mu_B}=\frac
{2(a^2+1)k|\int^\infty_0 uv\rho^3 d\rho|}{{3a}\int^\infty_0
(u^2+v^2)\rho^2 d\rho}-1.
  \label{3.17} \end{eqnarray}
The empirical value of the AMM of an electron is $\Dl g
=0.001159652$. The computational result suggests that (\ref{3.17})
might be the truth of the AMM.

\begin{figure}[ht] \centering
\includegraphics[width=12cm]{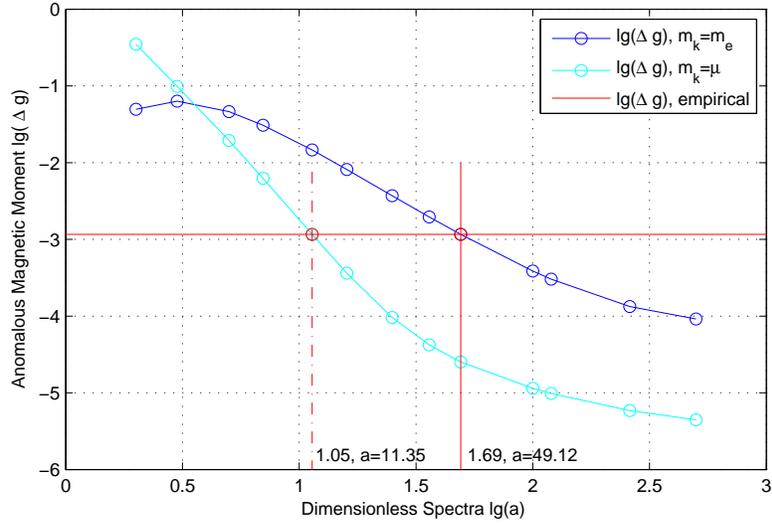}
\caption{The anomalous magnetic moment of the system (\ref{3.8})
vs. the spectra $a$, the true value for an electron is $\Dl g
=0.001159652$ or $\lg(\Dl g)=-2.936$} \label{fig4}
\end{figure}

To compare with the dark spinor, we also define the dimensionless
norm by (\ref{2.17}). The normalizing condition (\ref{2.16}) is
equivalent to $y=\lg S$. The dimensionless functions $(\E ,\Dl g
,y)$ are all continuous functions of $a$ for the same family of
solutions. Fig.\ref{fig4} shows how to determine the spectrum $a$
by the empirical AMM. Different definition of mass leads to
different value of $a$.
\begin{figure}[ht] \centering
\includegraphics[width=12cm]{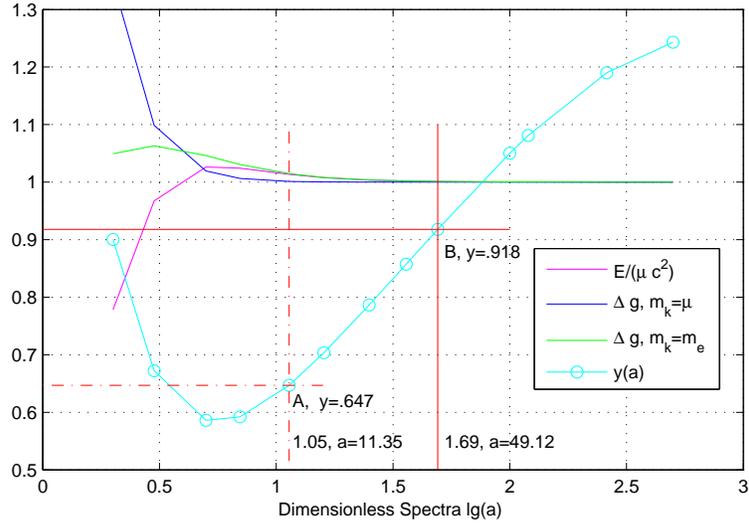}
\caption{The dimensionless functions $(\E(a),\Dl g(a),y(a))$. The
constant $S$ is determined by the intersection points A or B,
which correspond to the empirical anomalous magnetic moment}
\label{fig5}
\end{figure}

In Fig.\ref{fig4}, the trends of $\Dl g $ shows that $\Dl g $ is a
decreasing function of $a$, and $\Dl g \to 0$ as $a\to \infty$. By
the empirical data of $\Dl g$, we can compute the following
undetermined parameters, If taking $m_k=\E\mu$, we have
\begin{eqnarray}a=49.12,~ S=8.277,~ w=
3.385\times 10^{-57}{\rm Jsm}^2,~ E_{V}=1.088{\rm keV},~
E_{A}=85{\rm eV}. \label{3.21}\end{eqnarray} If taking $m_k=\mu$,
we have
\begin{eqnarray}a=11.35,~ S=4.434,~ w=
9.723\times 10^{-58}{\rm Jsm}^2,~ E_{V}=15.08{\rm keV},~
E_{A}=330{\rm eV}. \label{3.22}\end{eqnarray}

Fig.\ref{fig5} shows the realistic values of some parameters such
as the total energy $\E$, the norm function $y(a)$. The constants
$S$ or $w$ is determined by normalizing condition $y=\lg S$, and
then all other parameters can be computed. By Fig.\ref{fig5}, we
learn that, the value of $a$ is larger than that of dark spinor,
namely, the electromagnetic interaction increases the rest mass
$m$ of a spinor.
\begin{figure}[ht] \centering
\includegraphics[width=12cm]{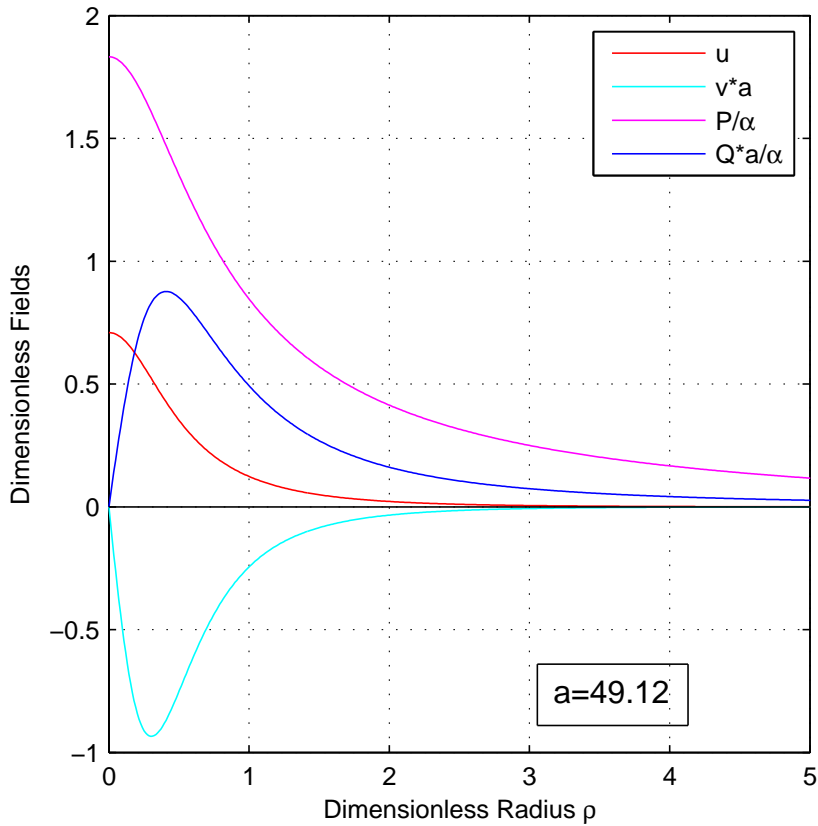}
\caption{The dimensionless radial functions, $(u,~v)$ correspond
to spinor fields. $(P ,~Q)$ correspond to the dimensionless
potentials.} \label{fig6}
\end{figure}

Since $\al\dot=\frac 1 {137}$ is quite small, by (\ref{*3.2}) we
learn that, if $a>10$, the electromagnetic field only have a
little influence on the eigen solution. The numerical results also
show this conclusion, Fig.\ref{fig6} shows the comparison of the
dimensionless fields when $a=49.12$.

\section{Some interesting properties of the nonlinear spinor}
\setcounter{equation}{0} Form the above results and some other
computation and analysis, we find some special but interesting
properties of the nonlinear spinor. These unusual properties might
have close relationship with the nature of the elementary
particles.
\begin{enumerate}

\item By $P\to 0$ and (\ref{2.20}), for $V=\frac 1 2 w \ck\ga^2$
we find
\begin{eqnarray}
mc^2\to \int^\infty_0 (\mu c^2 \ck\ga+\frac1 2 wc\ck\ga^2)\cdot
4\pi r^2 dr. \label{4.1}
\end{eqnarray}
More calculations show that such relation also hold for other kind
nonlinear potential $V(\ck\ga)$ satisfying $V'\ck\ga-V>0$, namely
we always have $|P|\ll E$. An interesting problem is whether the
error is just caused by numerical approximation and $P=0$ is a
rigorous relation generally valid for nonlinear spinors?

\item All dimensionless energy $\E(a)$ have a similar trend $\E\to
1(a\to \infty)$. For large enough $a$, we always have $E \to \mu
c^2$.

\item Taking equation system (\ref{3.8}) as a developing system
with respect to $\rho$, we find the initial value of $v(0),v'(0)$
etc. are determined by $u(0)$, then the eigen solutions satisfying
boundary condition (\ref{3.9}) only depend on $u(0)$. This is a
general feature for all eigen solutions of any nonlinear spinor,
which implies the eigen equation of the nonlinear spinor is
over-determined. This fact might be the underlying reason of the
Pauli principle\cite{eigen,commu}.

\item For the nonlinear spinor equation with a scalar interaction
\begin{eqnarray}
\al^\mu\hbar i\pa_\mu\phi = (\mu c+s G-V')\ga\phi,\qquad
(\pa_\al\pa^\al +b^2) G=\la s\ck\ga, \label{4.2}
\end{eqnarray}
similar to (\ref{3.10}) and (\ref{3.11}), $G$ can be expressed as
\begin{eqnarray}
G(r) = \frac {\la s} r \int^r_0 e^{-b(r-\tau)} d \tau
\int^\infty_\tau \ck\ga(\xi)\xi e^{-b(\xi-\tau)} d\xi,
\end{eqnarray}
so the solution to (\ref{4.2}) can be soundly solved by iteration.
For the AMM $\Dl g$ defined by (\ref{3.17}), computations show
that we always have $\Dl g\sim 0$ similar to the above cases with
electromagnetic interaction. This result implies that, it is
inadequate to describe the strong interaction by a scalar field.

\item Some rough calculations show that, the AMM of a proton might
be explained by the following nonlinear spinor with a strongly
coupling vector interaction $G^\mu$,
\begin{eqnarray}
\al^\mu(\hbar i\pa_\mu-e A_\mu+ s G_\mu )\phi = ({\rm M} c-V')
\ga\phi, \label{4.3} \\
\pa_\al\pa^\al A^\mu = e\ck\al^\mu, \quad (\pa_\al\pa^\al +b^2)
G^\mu = s\ck\al^\mu. \label{4.4}
\end{eqnarray}
If $\be=\frac{s^2}{4\pi\hbar} \sim 1$, we have the following
conclusions: (I). In the dimensionless form, the absolute values
of $(|G_0|,|\vec G|)$ are comparable with $|\phi|$ near the
center, so the first order approximation (\ref{3.1}) is invalid.
(II). The first family of the eigen solutions is absent, namely,
the fields $(g(r),f(r))$ have intersections with the horizontal
axis near the origin, so such spinor has complicated interior
structure. (III). Although the solution has $j_3  =\frac 1 2$, but
the solution $\phi$ includes components with orbital angular
momentum, which has strong influence on the AMM, so the value of
$|\Dl g|$ is much larger than that caused by the weakly coupled
electromagnetic interaction.

\item The energy functional of the nonlinear spinor system
(\ref{1.7}) is indefinite, so the stability of the solutions is
special. There are some works on this problem\cite{9,10,11}.
However, the usual treatment for the positive definite system
might be inadequate for the nonlinear spinors.

\item From the derivations of some previous works, we find that we
could almost reconstruct the physical theories based on the
nonlinear spinors, so some further investigations on the nonlinear
spinors are worthwhile.

\end{enumerate}

\section{The Test of Mass-Energy Relation}
\setcounter{equation}{0}

The Einstein's mass-energy relation $E=mc^2$ is one of the most
fundamental formulae in physics, but it has not been seriously
tested by an elaborated experiment, and only some indirect
evidences in nuclear reaction suggested that it holds to high
precision. From the above calculation, we found the interaction
potentials of a particle will yield detectable effects, which lead
to different energy-speed relation, which can be used as the
fingerprints of the interactive potentials of elementary
particles. So the experiment may shed lights on the nature of the
interaction and elementary particles.

In what follows, we give a detailed description for the
experiment. The experiment only involves low energy accelerator of
particles and measurement of speed. In this section, $(u,v)$ stand
for the speed of a particle, which can not be confused with the
fields defined in (\ref{1.12}) or (\ref{2.13}).

In Einstein's original paper \cite{TE1}, he derived the kinetic
energy of a particle $K$,
\begin{eqnarray}K=mc^2-m_0 c^2,
\qquad m={m_0}\l(1-\frac {v^2}{c^2}\r)^{-\frac 1 2},
\label{T1.1}\end{eqnarray} which implies the total energy and the
speed of a particle have the following simple relation
\begin{eqnarray}E=m(v) c^2. \label{T1.2}\end{eqnarray} However,
this relation is based on the linear classical mechanics, and it
has not been directly tested by elaborated experiment. There were
once some indirect evidences in the nuclear reaction. The most
accurate one is provided by S. Rainville {\it et al}\cite{Ttest1},
which indicates that the mass-energy relation $E=m c^2$ holds to
an error level less than 0.00004\% in the process of neutron
capture by nuclei of sulfur and silicon resulting in
$\gamma$-radiation. As pointed out by E. Bakhoum\cite{Ttest2},
although the authors claimed it is a direct test, it is actually a
test for the energy conversion $\Dl E=\Dl m c^2$ at low speed of
the particles. As one of the most fundamental relation, a direct
test for the original energy-speed relation (\ref{T1.2}) is
necessary and significant.

What more important is that, for the nonlinear spinors with
interactive potentials such as electromagnetic one $A^\mu$, The
detailed calculation shows the interaction terms result in the
fine structure of the energy-speed relation\cite{Tgu1,Tgu2}, and
the fine structure can be used as fingerprints of the
interactions. (\ref{3.21}) and (\ref{3.22}) show that, the
nonlinear potential yields energy to a magnitude of 1keV, and the
electromagnetic interaction to a magnitude of 100eV, which can be
easily detected by elaborated experiments.

Hereafter we take $c=1$ as unit of speed. The general
representation of the energy-speed relation is given by
\begin{eqnarray}
E(u)= \frac {M_0}{\sqrt{1-u^2}} -\frac{M_1
u^2}{\sqrt{1-u^2}}+\frac{M_\ga}{\sqrt{1-u^2}}
\ln\frac{1}{\sqrt{1-u^2}},\label{T1.3}
\end{eqnarray}
where $(M_0,M_1,M_\ga)$ are all constants of mass dimension, and
$M_0$ is the total static mass of the particles, $M_1$ corresponds
to interactions such as electromagnetic potential, $M_\ga$
corresponds to the nonlinear self-interactive potential.  The
detailed explanation of the parameters see \cite{Tgu2}.

Because of the little value of $(M_1,M_\ga)$ and the function
$\ln\sqrt{1-u^2}$, the nonlinear effects can be easily concealed
behind $M_0$. For example, when an electron get kinetic energy
30MeV. The corresponding speed reaches $u_1=0.99986c$, but
\begin{eqnarray}
E(u_1)\dot=\frac{M_0-M_1+4 M_\ga}{\sqrt{1-u_1^2}}\approx
\frac{M_0}{\sqrt{1-u_1^2}}.\label{T1.6*}
\end{eqnarray}
That is to say, (\ref{T1.3}) is a stiff equation of the
coefficients $(M_0,M_1,M_\ga)$. So we have to make some
transformation to get meaningful solution.

\begin{figure}
\centering
\includegraphics[width=16cm]{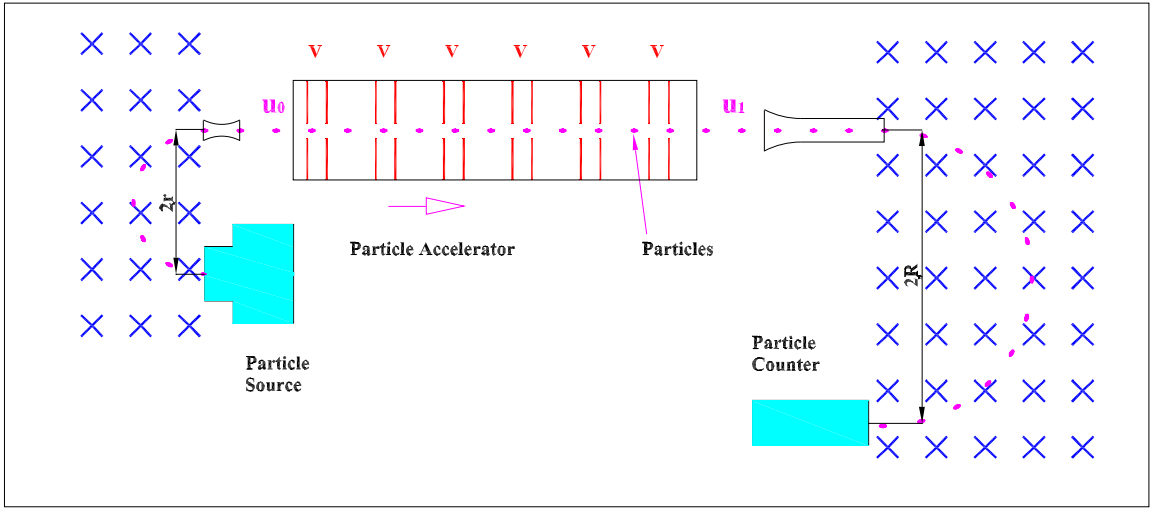}
\caption{The flow chart and experimental scheme to test the
mass-energy relation} \label{Tfig}
\end{figure}

We propose the following experimental project to measure the
coefficients $(M_0,M_1,M_\ga)$ in the energy-speed relation. The
flow chart and experimental scheme are illustrated in
Fig.\ref{Tfig}. The particles with unit charge are produced by the
particles source, and the ones at given initial speed $u_0$ are
selected by a homogeneous magnetic field. By adjusting the radius
$r$, we can control the initial speed of the particles $u_0$. The
series-wound accelerator is constructed by a set of uniform
electrodes, which can be charged with high voltage $V$. When the
selected particles pass though one pair electrodes, each particle
receives an energy increment $\dl E=eV$, which converts into its
kinetic energy. If $n$ pair electrodes are charged, then we get
the total kinetic energy increment for each particle
\begin{eqnarray}
K=E(u_1)-E(u_0) = n\dl E =neV.\label{T1.4}
\end{eqnarray}
(\ref{T1.3}) and (\ref{T1.4}) establish the connection between the
speed $u$ and $nV$. The final speed $u=u_1$ of the particles can
be measured by the position $R$ of the particle counter or film.
Then we can determine the constants $(M_0,M_1,M_\ga)$ by fitting
the curve $f(u_1,V)=0$ as defined by (\ref{T1.4}).

Now we make some simplification of (\ref{T1.4}). At first, we can
solve the static mass $M_0$ at low energy $u=\wt u_1\ll c$ as
follows. Assume the voltage $V=V_0$, in this nonrelativistic case,
we have the approximation of (\ref{T1.4}) as follows
\begin{eqnarray}
neV_0 &\dot = &\frac 1 2 (M_0-2 M_1+M_\ga)(\wt
u_1^2-u_0^2).\label{T1.5}
\end{eqnarray}
Then we get
\begin{eqnarray}\l\{\begin{array}{lll}
M_0&=&m_s+2 M_1-M_\ga, \\
E_0&=& m_s+2M_1-M_\ga+\frac 1 2 m_su_0^2\dot =E(u_0),\\
m_s&\equiv&2n e V_0(\wt u_1^2-u_0^2)^{-1},\end{array}\r.
\label{T1.6}
\end{eqnarray}
where $m_s$ is the non-relativistic static mass of the particle in
classical sense. Therefore, we only need to determine two little
coefficients $(M_1,M_\ga)$ at high energy.

By (\ref{T1.3}), (\ref{T1.4}) and (\ref{T1.6}), we have the
following relation
\begin{eqnarray}
(2-u_1^2)M_1&-&\l(1+\ln\sqrt{1-u_1^2}\r)M_\ga  \l\{=
E(u_1)\sqrt{1-u_1^2}-m_s, \r.
\qquad {\rm (by ~(\ref{T1.3})~ and~ (\ref{T1.6}))}\nn\\
& = & \l. [neV+E(u_0)]\sqrt{1-u_1^2}-m_s \r\}, \qquad\qquad
\qquad\qquad\quad~
{\rm (by~ (\ref{T1.4})~ and~ (\ref{T1.6}))}\nn\\
& \dot = & (neV+\frac 1 2 m_s u_0^2)\sqrt{1-u_1^2}-\frac {m_s
u_1^2}{1+\sqrt{1-u_1^2}}+(2M_1-M_\ga)\sqrt{1-u_1^2}. \label{T1.7}
\end{eqnarray}
Denoting
\begin{eqnarray}
\chi  = \frac {u_1^2}{1+\sqrt{1-u_1^2}},\qquad U=(neV+\frac 1 2
m_s u_0^2)c^{-2}, \label{T1.7*}
\end{eqnarray}
and substituting it into (\ref{T1.7}), we get
\begin{eqnarray}
\chi ^2 M_1-\l[\chi +\ln({1-\chi })\r]M_\ga = ({1-\chi })U - \chi
m_s.\label{T1.8}
\end{eqnarray}
(\ref{T1.8}) is a linear equation of $(M_1,M_\ga)$, which can be
easily solved by the method of least squares from a sequence of
measured data $(U_i, \chi_i),i=1,2,\cdots,N$. Define the
$N$-dimensional vectors  $(\vec a, \vec b, \vec f)$ by,
\begin{eqnarray}
\vec a&=& (\chi_1^2,\chi_2^2,\cdots, \chi_N^2),\\
\vec b&=&
-\l[\chi_1+\ln({1-\chi_1}),\chi_2+\ln({1-\chi_2}),\cdots,
\chi_N+\ln({1-\chi_N})\r],\\
\vec f &=&[ ({1-\chi_1})U_1 -\chi_1 m_s, ({1-\chi_2})U_2 -\chi_2
m_s,\cdots,({1-\chi_N})U_N -\chi_N m_s].
\end{eqnarray}
Then the solution of the least square is given by
\begin{eqnarray}
M_1=\frac 1 D [\vec b~^2\vec a-(\vec a\cdot\vec b)\vec b]\cdot
\vec f,\quad M_\ga=\frac 1 D [\vec a^2\vec b-(\vec a\cdot\vec
b)\vec a]\cdot \vec f,\quad D = \vec b~^2 \vec a^2 -(\vec
a\cdot\vec b)^2,
\end{eqnarray}

From the above computation, for an electron we have the typical
order of magnitude for the parameters in (\ref{T1.8}),
\begin{eqnarray}
\chi \sim 1,\quad U\sim m_s\sim 1{\rm MeV},\quad M_1\sim 100 {\rm
eV},\quad M_\ga\sim 1 {\rm keV}.
\end{eqnarray}
The synchrotron radiation is much less than $M_1$ and $M_\ga$, so
it can be omitted. A meaningful test strongly depends on the
precision of the measurement data $(\chi_i,U_i)$, which should be
of relative errors less than $10^{-3}$. How to promote the
precision of the measurement is the key for the success of a test.

Some possible solutions and its implications:
\begin{enumerate}
\item If $M_\ga=0$ and $M_1=0$, which means the Einstein's
mass-energy strictly holds, and the particles can not be described
by the classical fields. \item If $M_\ga=0$ and $M_1\ne 0$, this
kind of particles has not nonlinear self-interaction, and the
balance of the particles should be explained by scalar and vector
interactions. \item If $M_\ga \ne 0$, which means the standard
model of particles is incomplete, and some calculations in quantum
field theory should be modified.
\end{enumerate} So no matter what result the experiment provides,
the implication is always important and fundamental. Although the
standard model of particles has achieved a lot of progresses in
explanation of the behavior of micro particles, it is essentially
a phenomenology. So the test of the energy-speed relation might be
a shortcut to disclose the secrets of the fundamental particles
and their interactions.

\section{Discussion and Conclusion}
\setcounter{equation}{0}

we have solved the particle-like eigen solutions to some nonlinear
spinor equations, and computed several functions which reflect
their characteristics. The numerical results show that, the
nonlinear spinor equations have positive discrete mass spectra and
anomalous magnetic moment. The nonlinear potential and
interactions yield different contributions to the total energy of
the system, which can be used as fingerprints of these terms. The
magnitude of the components can be easily detected by elaborate
experiments. The weird properties of the nonlinear spinors might
be closely related with the elementary particles and their
interactions, so some deeper investigations on them are
significant.

\section*{Acknowledgments}
The author is grateful to Prof. Ta-Tsien Li, Prof. Han-Ji Shang
and Prof. Guang-Jiong Ni for their guidance. Thanks to Prof.
Yue-Liang Wu and Dr. Ci Zhuang of the Chinese Academy of Science
for encouragement of this program.

\end{document}